\documentclass{article}
\usepackage{spconf,amsmath, amssymb, graphicx}
\usepackage{multirow}
\usepackage{booktabs}
\usepackage[textfont=it,tableposition=top]{caption}
\usepackage{color}
\usepackage{soul}
\usepackage{float}

\def\x{{\mathbf x}}
\def\y{{\mathbf y}}
\def\yhat{{\mathbf{\hat{y}}}}
\def\henc{{\mathbf h}^\text{enc}}

\def\blank{\left<\text{b}\right>}
\def\sos{\left<\text{sos}\right>}

\title{Recognizing Long-Form Speech Using Streaming End-to-End Models}
\name{Arun Narayanan, Rohit Prabhavalkar, Chung-Cheng Chiu}
\secondlinename{David Rybach, Tara N. Sainath, Trevor Strohman}
\address{Google, USA}

\begin{document}
\ninept
\maketitle
\begin{abstract}
All-neural end-to-end (E2E) automatic speech recognition (ASR) systems that use a single neural network to transduce audio to word sequences have been shown to achieve state-of-the-art results on several tasks.
In this work, we examine the ability of E2E models to generalize to unseen domains, where we find that 
models trained on short utterances fail to generalize to long-form speech.
We propose two complementary solutions to address this: training on diverse acoustic data, and LSTM state manipulation to simulate long-form audio when training using short utterances. 
On a synthesized long-form test set, adding data diversity improves word error rate (WER) by 90\% relative, while simulating long-form training improves it by 67\% relative, though the combination doesn't improve over data diversity alone.
On a real long-form call-center test set, adding data diversity improves WER by 40\% relative. Simulating long-form training on top of data diversity improves performance by an additional 27\% relative. 
\end{abstract}
\section{Introduction}
\label{sec:introduction}
Automatic speech recognition (ASR) enabled technologies have become increasingly prominent and mainstream over the last few years. 
From mobile phones, to smart digital assistants (e.g., Google Home, Amazon Alexa, Apple Siri, and Microsoft Cortana), ASR technologies power a wide variety of user interactions.
While this widespread adoption owes its success to deep learning~\cite{HintonDengYuDahlEtAl12}, the technology at the heart of state-of-the-art ASR systems has remained fairly static over the last couple of decades; context-dependent phonemes as acoustic units and weighted finite state transducers (WFSTs) remain integral parts of a conventional ASR system~\cite{MorganBourlard95}.

The dominance of conventional hybrid speech recognizers, however, is being challenged by the growing body of research which has focused on end-to-end (E2E) models. 
Such models replace the components of a traditional ASR system -- the acoustic model (AM), the pronunciation model (PM), and the language model (LM) -- with a single, all-neural network that directly produces output character or word units given input acoustic frames.
Examples include connectionist temporal classification (CTC) with character~\cite{HannunCaseCasperCatanzaroEtAl14, AmodeiAnathanarayananAnubhaiBaiEtAl16} or word-based~\cite{soltau2016neural, AudhkhasiRamabhadranSaonPichenyEtAl17} targets, the recurrent neural network transducer (RNN-T)~\cite{Graves12, GravesMohamedHinton13}, and attention-based encoder-decoder models \cite{ChorowskiBahdanauSerdyukChoEtAl15, ChanJaitlyLeVinyals16, PrabhavalkarRaoSainathLiEtAl17, BattenbergChenChildCoatesEtAl17, HoriWatanabeZhangChan2017, ZeyerIrieSchluterNey2018}.
On medium-scale tasks such as Librispeech~\cite{ParkChanZhangChiuEtAl19} or large-scale voice-search tasks~\cite{RaoSakPrabhavalkar17, ChiuSainathWuPrabhavalkarEtAl18, TaraRuoming19}, such E2E models have been shown to outperform conventional ASR systems.

The acoustic and linguistic characteristics of speech vary widely across \emph{domains}.\footnote{We use the term \emph{domain} loosely in this context to refer to a group of utterances that share some characteristics. Examples of domains might include `telephony' (i.e., spontaneous speech sampled at 8kHz), `audiobooks' (i.e., long utterances of read speech), etc.}
Ideally, we would like to train a single ASR system that performs equally well across all domains~\cite{NarayananMisraSimPundakEtAl18}.
Building an E2E ASR system that generalizes well to multiple domains is particularly challenging since E2E models learn all system components directly from the training data.
In this work, we investigate the generalization ability of \emph{streaming} E2E ASR systems -- specifically, the recurrent neural network transducer (RNN-T)~\cite{Graves12, GravesMohamedHinton13}\footnote{In the authors' experience, RNN-T models are better suited for streaming long-form ASR tasks, compared to other E2E variants \cite{jaitly2015neural,ChiuRaffel18, Chiu19longform}.} -- to determine its robustness to domain mismatches between training and testing.
We find that the tight coupling 
results in a large performance degradation due to such a domain mismatch.
We also identify a related, but distinct limitation of such E2E models -- their inability to decode \emph{long-form speech} (e.g., long YouTube videos), when trained on short training utterances.
We consider two complementary solutions that are targeted towards alleviating these issues: training on \emph{diverse acoustic domains}, which improves generalization accuracy on unseen domains; and \emph{simulating long-form training} by manipulating the model's internal recurrent state, which improves accuracy on long-form speech.
On an unseen domain of real long-form utterances comprised of anonymized call-center conversations, we find that training with diverse data improves word error rate (WER) by 40\%; simulating long-form training data provides an additional 27\% WER reduction without significantly degrading performance on any of the in-domain test sets.

The structure of the rest of this paper is as follows: In Sec.~\ref{sec:rnnt}, we describe the RNN-T model used in this work.
In Sec.~\ref{sec:improving-longform}, we describe the specific techniques proposed to improve performance on out-of-domain data: increasing training data diversity (Sec.~\ref{sec:data-diversity}) and simulating long-form speech during training (Sec.~\ref{sec:simulting-long-form}).
We describe our experimental setup and the training and test sets used in this work in Sec.~\ref{sec:experiments}, and report results and analysis in Sec.~\ref{sec:results} and~\ref{sec:analysis}, respectively, 
before concluding in Sec.~\ref{sec:conclusions}. 

\section{Recurrent Neural Network transducer}
\label{sec:rnnt}
\begin{figure}
  \centering
  \includegraphics[width=0.9\columnwidth]{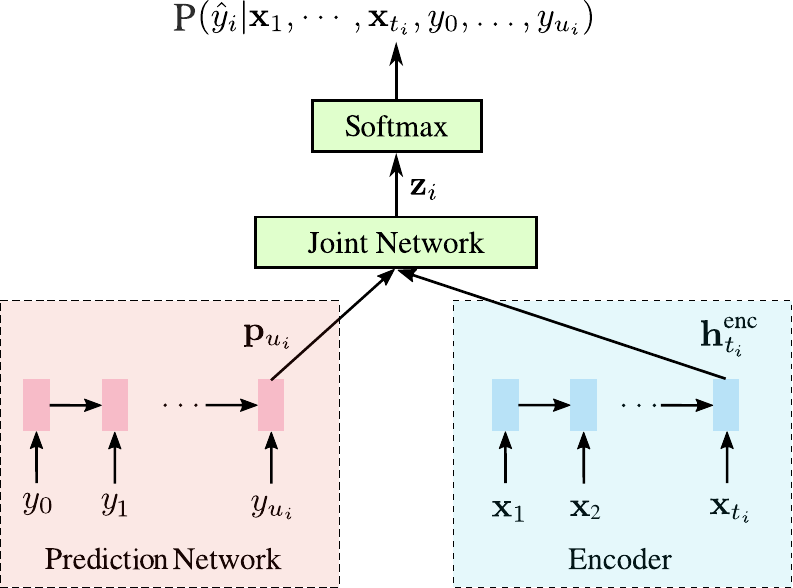}
  \caption{A schematic representation of the RNN-T model.}
  \label{fig:rnnt}
\end{figure}
We denote the input speech utterance as $\x = [\x_1, \cdots, \x_T]$, where $\x_t \in \mathbb{R}^d$ correspond to individual acoustic frames. During training, we assume that we are also provided with the corresponding label sequence, $\y = [y_1, \cdots, y_U]$, where $y_u \in \mathcal{Y}$, the set of output labels. We use word-pieces~\cite{SchusterNakajima12} as output units in this work\footnote{For example, \texttt{the cat basks in the sun} might be decomposed into the word-pieces \texttt{\_the \_cat \_b ask s \_in \_the \_sun}.}.

The RNN-T model was proposed by Graves~\cite{Graves12, GravesMohamedHinton13}, as an extension of CTC~\cite{GravesFernandezGomezSchmidhuber06}, and is depicted in Fig.~\ref{fig:rnnt}.
The model defines a probability distribution over the set of output labels augmented with a special blank symbol,  $\blank$. 
The blank symbol is introduced to account for the fact that we do not have frame-level alignments for the labels.
We define the set of all valid frame alignments, $B(\x, \y)$, as the set of all label sequences, $\yhat = (\hat{y}_1, \cdots, \hat{y}_{T+U})$, where $\hat{y}_i \in \mathcal{Y} \cup \left\{\blank\right\}$, such that $\yhat$ is \emph{identical} to $\y$ after removing all blank symbols.
Thus, by construction, all sequences in $B(\x, \y)$ contain exactly $T$ blank symbols.

As can be seen in Fig.~\ref{fig:rnnt}, the RNN-T model consists of three main components: the \emph{encoder}, the \emph{prediction network} (both modeled using LSTMs~\cite{HochreiterSchmidhuber97}), and the \emph{joint network} (modeled using a feed-forward network).
The RNN-T model defines the total probability of the output sequence by marginalizing over all possible alignments:
\begin{align}
P(\y | \x) &= \sum_{\yhat \in B(\x, \y)} P(\yhat|\x) \\
           &= \sum_{\yhat \in B(\x, \y)} \prod_{i=1}^{T+U} P(\hat{y}_i | \x_1, \cdots, \x_{t_i}, y_0, \cdots, y_{u_i}) \label{eqn:rnnt}
\end{align}
\noindent where, $y_0=\sos$ is a special symbol denoting the start of the utterance; $u_i$, and $t_i - 1$ are the total number of non-blank, and blank symbols respectively in the partial alignment sequence $(\hat{y}_1, \cdots, \hat{y}_{i-1})$.
Thus, the outputs at every frame are conditioned on the sequence of previous predictions.

Specifically, the encoder (analogous to the AM in a traditional ASR system) receives an input acoustic frame, $\x_{t_i}$, before transforming it into a higher-level representation, $\henc_{t_i}$.
The prediction network (analogous to an RNN LM) receives as input the previous \emph{non-blank} label, $y_{u_i} \in \mathcal{Y}$, and computes an embedding, $\mathbf{p}_{u_i}$, which is fed to the joint layer. 
Finally, the joint network computes logits, $\mathbf{z_i}$, by combining the outputs of the two other layers through two feed-forward layers -- first with a $\tanh$ activation function, followed by a linear activation function.
When the output from the final softmax layer corresponds to the blank symbol, the encoder is fed with the next input frame while leaving the prediction network state unchanged; if the output from the final softmax layer is \emph{non-blank}, the label is fed to the prediction network while leaving the encoder state unchanged.

The RNN-T model can be trained by optimizing log-likelihood over the training set: $\mathcal{L} = \sum_{(\x, \y)} \log P(\y|\x)$, and the required gradients can be computed using dynamic programming~\cite{Graves12, GravesMohamedHinton13}. 

\subsection{Inference}
Inference in the model is performed using beam search~\cite{SutskeverVinyalsLe14}. 
In order to ensure that recognition is streaming, we use a `frame-synchronous' search strategy.
Specifically, given a set of beam candidates, we first feed the new input frame to the encoder. 
All hypotheses are then repeatedly expanded until the most-likely next symbol is the blank symbol (at which point the hypothesis is a candidate for the beam at the next frame).
We also employ an adaptive beam \cite{Lowerre76}, which prunes partial hypotheses which are much worse than the current best hypothesis.
Finally, we merge two beam candidates $\yhat$ and $\yhat'$ during the search, if they correspond to different alignments of the same sequence, by summing up their probabilities, which we find significantly improves performance\footnote{E.g., \texttt{$\yhat=$~\_the $\blank$ \_cat $\blank$} and \texttt{$\yhat'=$~$\blank$ \_the \_cat $\blank$}.}.

\section{Improving long-form modeling}
\label{sec:improving-longform}

As described in Sec.~\ref{sec:rnnt}, E2E ASR systems directly model the posterior of word sequences given audio.
Such models do not require frame-level alignments; instead, alignments are learned implicitly during training, with recurrent neural networks such as LSTMs~\cite{HochreiterSchmidhuber97}.
Consequently, the models tend to be more sensitive to the specific characteristics of the acoustics they are trained with compared to non-recurrent models, or models that use pre-computed alignments.
As we show in this work, if the acoustic characteristics of speech differ during inference, the model's performance can degrade significantly.
For example, a model trained on anonymized voice-search utterances performs poorly when captioning YouTube videos; the degradation is more severe than would be expected in conventional ASR systems~\cite{NarayananMisraSimPundakEtAl18}.

A related, but more subtle issue, is the lack of generalization to long audio sequences.
As we show in Sec.~\ref{sec:results}, decoding a long-duration audio test set with a model trained using short-duration voice-search queries results in poor performance.
Part of the difficulty in decoding long audio sequences comes from the fact that this condition differs significantly from how models are trained.
Since hardware accelerators typically have small amounts of on-device memory, it is challenging to use extremely long utterances with mini-batch training.
Therefore, training utterances are segmented into smaller chunks -- typically a few seconds long.
Moreover, training E2E models, which have recurrent structures, on long utterances is challenging because of the vanishing gradient problem~\cite{hochreiter1998vanishing}. 
The problem is exacerbated because the model has to implicitly align long segments when learning $P(\y|\x)$.
In the following sections, we discuss alternatives to training directly on long-utterances.

\subsection{Increasing Data Diversity Through Multidomain Training}
\label{sec:data-diversity}
Given the large capacity of neural net based models, specific mismatches are most commonly addressed by collecting a wide-variety of training data. 
Techniques such as multi-condition training have been shown to improve performance in noisy and farfield conditions~\cite{peddinti2016mtr,kim2017mtr}.
To address domain mismatch, multidomain training has been proposed with considerable success for conventional neural network acoustic models~\cite{NarayananMisraSimPundakEtAl18,ghahremani17-transfer}.
We explore a similar strategy of expanding training data to include a diverse set of conditions. Apart from addressing acoustic mismatches, multidomain data also increases the diversity of utterance lengths.

\subsection{Simulating Long-Form Characteristics During Training}
\label{sec:simulting-long-form}
As mentioned before, the constraints imposed by modern hardware accelerators restrict our ability to train on very long audio segments.
Since the model's recurrent state (i.e., the state of the LSTMs) evolves over time, this introduces a mismatch between training and inference.
During training, the LSTM states are reset after processing each training utterance; during inference on long-form audio, however, the LSTM states are propagated without reset for several time-steps. 
Since the LSTM states convey almost all\footnote{If there is an explicit conditioning on previous word tokens in the model, as is the case with most E2E models, apart from LSTM states, the previous output tokens also have to be propagated.} the information that is to be propagated from one step to the next, we consider algorithms that manipulate the \emph{initial} LSTM states. 

Let $L(0)$ represent the initial state of the model. 
For RNN-T, $L(0) \equiv \{L_{e}(0)$, $L_{p}(0)\}$, where $L_{e}(0)$ and $L_{p}(0)$ are the initial states of the encoder and the prediction network's LSTM layers, respectively.
Specifically, for LSTMs, $L_{i}(0) \equiv [C_{i}(0), M_{i}(0)]$, for $i \in \{e, p\}$, where $C_{i}(0)$ and $M_{i}(0)$ are the initial cell state and memory, respectively.
Typically, $C_{i}(0)$ and $M_{i}(0)$ are set to all-0 vectors, $\mathbf{0}$.
The states evolve in a recurrent fashion:
\begin{equation}
\begin{split}
L^{k}(0) &= \mathbf{0}, \\
L^{k}(t, u) &= \text{LSTM}(\x^{k}_t, y^{k}_u, L^{k}(t-1, u-1)),\\
& \quad\quad\quad\forall t, u \in [1, \ldots, T^{k}; 1, \ldots, U^{k}].
\end{split}
\end{equation}
Here, $T^k$ is the total number of frames, $U^k$ is the number of output tokens, and $k$ indexes training examples. We omit the index $k$, when it is clear from context.

\subsubsection{Random State Sampling (RSS)}
\label{sec:random-state-sampling}
The first technique that we investigate in the context of manipulating initial LSTM states is random state sampling (RSS).
In this technique, we assume that LSTM states follow a certain distribution. 
To simulate long segments of audio, instead of setting initial LSTM states to an all-0 vector, as is typically done during training, we sample initial states from the assumed distribution.
In this work, for simplicity, we assume that the LSTM states follow a random normal distribution:
\begin{equation}
L(0) \sim \mathcal{N}(\mathbf{0}, \mathcal{I})
\end{equation}
During inference, we initialize LSTM states to all-0 vector -- the expected value of the initial states used during training. 

\subsubsection{Random State Passing (RSP)}
\label{sec:random-state-passing}
\begin{figure}[t]
  \centering
  \includegraphics[width=0.9\columnwidth]{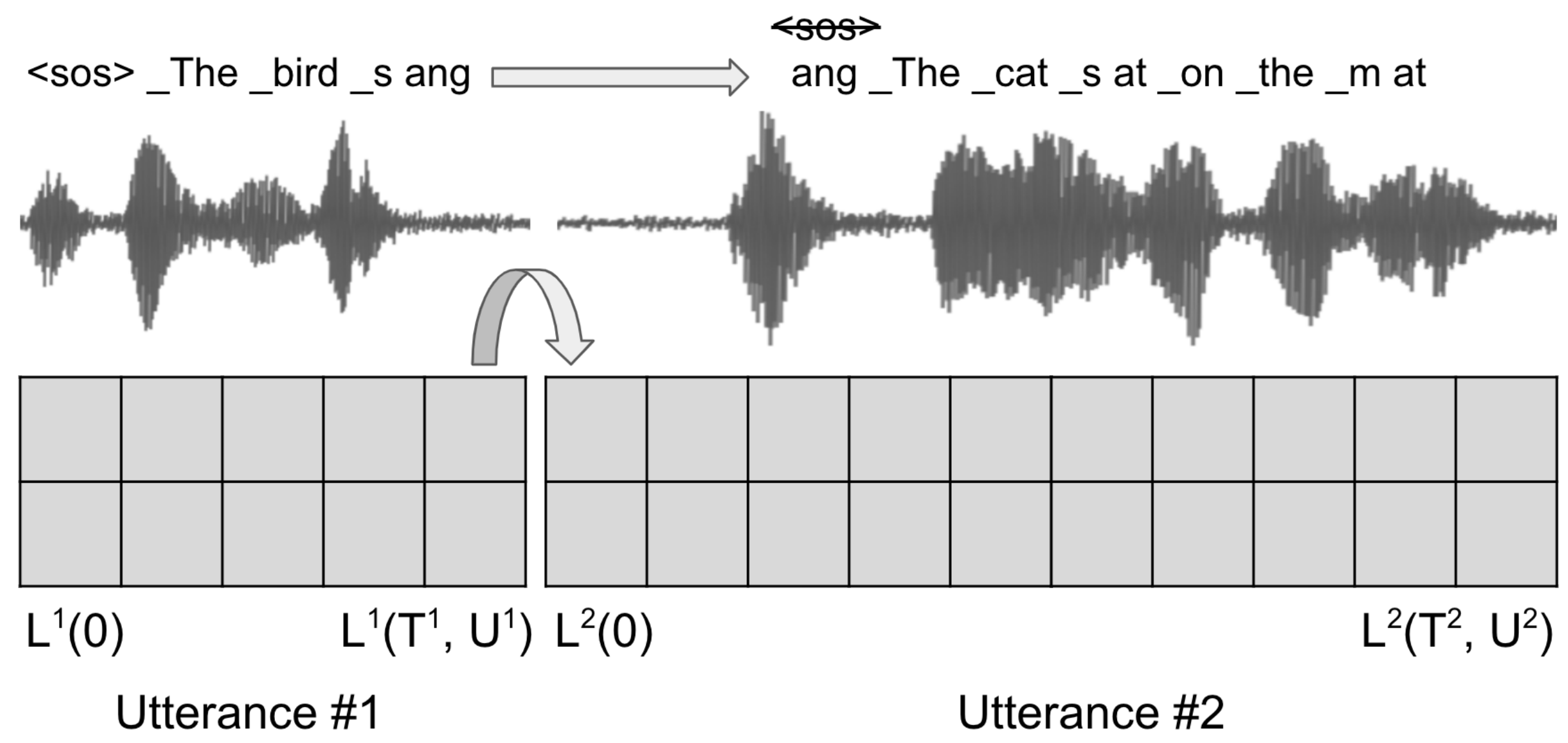}
  \caption{An example that illustrates random state passing (RSP). We can simulate concatenation of two utterances by using the last LSTM state of the first utterance, $L^{1}(T^{1}, U^{1})$, as the initial state of the second utterance, $L^{2}(0)$, and by replacing the start-of-speech token ($\sos$) for the second utterance by the last output token of the first utterance.}
  \label{fig:rsp}
\end{figure}
The second technique that we investigate, random state passing (RSP), simulates long-form audio more directly.
Instead of sampling from an assumed distribution, we save final states from every mini-batch as we train. 
The initial states for subsequent examples are then sampled from these saved states. 
In addition to the LSTM states, we also save and propagate the final output token from the utterances in the mini-batch. 
Thus, the technique is equivalent to concatenating training utterances and their transcripts.
The process is illustrated in Fig.~\ref{fig:rsp}.

While we propagate the states forward, the gradients are not passed back across mini-batches.
This is similar in spirit to the truncated back propagation through time algorithm, which is widely used for training recurrent neural nets~\cite{Werbos88}.
The idea of propagating just the states forward has recently also been used for training transformer-XL language models to learn long context~\cite{DaiYangYangCarbonnelEtAl19}.
In the current work, we propagate states across utterances with probability 0.5, and reset the LSTM states to all-zero vectors otherwise:
\begin{equation}
\begin{split}
L^{k}(0) &= 
  \begin{cases}
    \mathbf{0}, &~\text{with probability}~0.5 \\
    L^{j}_{e}(T^{j}),L^{j}_{p}(U^{j}) &~\text{with probability}~0.5 
 \end{cases} \\
L^{k}(t, u) &= \text{LSTM}(\x^{k}_t, y^{k}_u, L^{k}(t-1, u-1)),\\
& \quad\quad\quad\forall t, u \in [1, \ldots, T^{k}; 1, \ldots, U^{k}],
\end{split}
\end{equation}
\noindent where, $j$ corresponds to an utterance in a different mini-batch.
$L^{k}(T^{k}, U^{k})$ is saved to be used as the initial state for subsequent examples. 
During inference, we use an all-0 initial state for the LSTMs. 

\section{Experimental settings}
\label{sec:experiments}
We conduct detailed experiments on various training sets to determine the effectiveness of the proposed methods. Our experimental setup, training and evaluation sets are described in the subsequent sections.

\subsection{Model Architecture}
\label{subsec:archi}
As our acoustic frontend, we use 128-dimensional log-mel filterbank energies computed on {32~msec} windows with a {10~msec} hop.
Features from 4 contiguous frames are stacked, and then sub-sampled by a factor of 3.
Thus, the RNN-T model receives inputs at a frame rate of {30~msec}.
All experiments use multi-condition training for data augmentation and to simulate noisy conditions~\cite{kim2017mtr}.
Additionally, we generate both {8~kHz} and {16~kHz} versions of the training data to reduce acoustic mismatch due to sampling rates~\cite{NarayananMisraSimPundakEtAl18,Yu2013FeatureLearningDNN}.

Our RNN-T model architecture is similar to the architecture proposed in~\cite{HeSainathPrabhavalkarMcGrawEtAl19}: The model uses 8 unidirectional LSTM~\cite{HochreiterSchmidhuber97} layers for the encoder and 2 unidirectional LSTM layers for the prediction network. 
The encoder and the prediction network LSTMs have $2,048$ units each, which are projected down to $640$ output units after each layer. 
The encoder network also uses a time-reduction layer after the $2^\text{nd}$ LSTM layer, which stacks output features from 2 contiguous timesteps and subsamples them by a factor of 2;
the final encoder outputs are thus at a {60~msec} frame rate.
The joint network has a single layer with 640 units. 
The target transcript is represented as a sequence of word-piece tokens~\cite{SchusterNakajima12} with a vocabulary size of $4,096$; these are embedded into 128 dimensions before being fed into the prediction network.

In total, the models contain 120 million trainable parameters.
They are trained using TensorFlow~\cite{abadi2016tensorflow} and the open-source Lingvo toolkit~\cite{ShenNguyenWuChenEtAl19}, using $8 \times 8$ Tensor Processing Units (TPUs)~\cite{jouppi2017datacenter}.
For optimization, we use mini-batches of size $4,096$ with an Adam optimizer~\cite{kingma2014adam} and synchronized stochastic gradient descent.

\subsection{Datasets}
\label{subsec:datasets}
To evaluate the proposed techniques, we use a variety of training and test sets, all containing English speech.
The training sets include data from four domains: anonymized and hand-transcribed utterances representative of voice-search and far-field use cases, segmented telephony speech, and semi-supervised YouTube video segments~\cite{liao2013large,soltau2016neural}.
The amount of audio in the training set for each of these domains is shown in Tab.~\ref{tab:train_data}.
We report results obtained by training on just the voice-search subset (Search), or on data from all domains (multidomain).
\begin{table}[ht]
  \caption{Utterance length statistics for training data in this work.}
  \label{tab:train_data}
  \centering
  \begin{tabular}{crrr}
    \toprule
    \textbf{Application} & \textbf{Total} & \textbf{Mean} & \textbf{Median} \\
    \textbf{Domain} & \textbf{(hours)} & \textbf{(sec.)} & \textbf{(sec.)}  \\
    \midrule
    Search    & $56$k  & $6.2$ & $4.8$ \\
    Farfield  & $38$k  & $3.9$ & $3.5$ \\
    Telephony & $4$k   & $4.4$ & $3.0$ \\
    YouTube   & $190$k & $5.9$ & $4.5$ \\
    \bottomrule
  \end{tabular}
\end{table}
\begin{figure}[ht]
  \centering
  \includegraphics[width=1.0\linewidth]{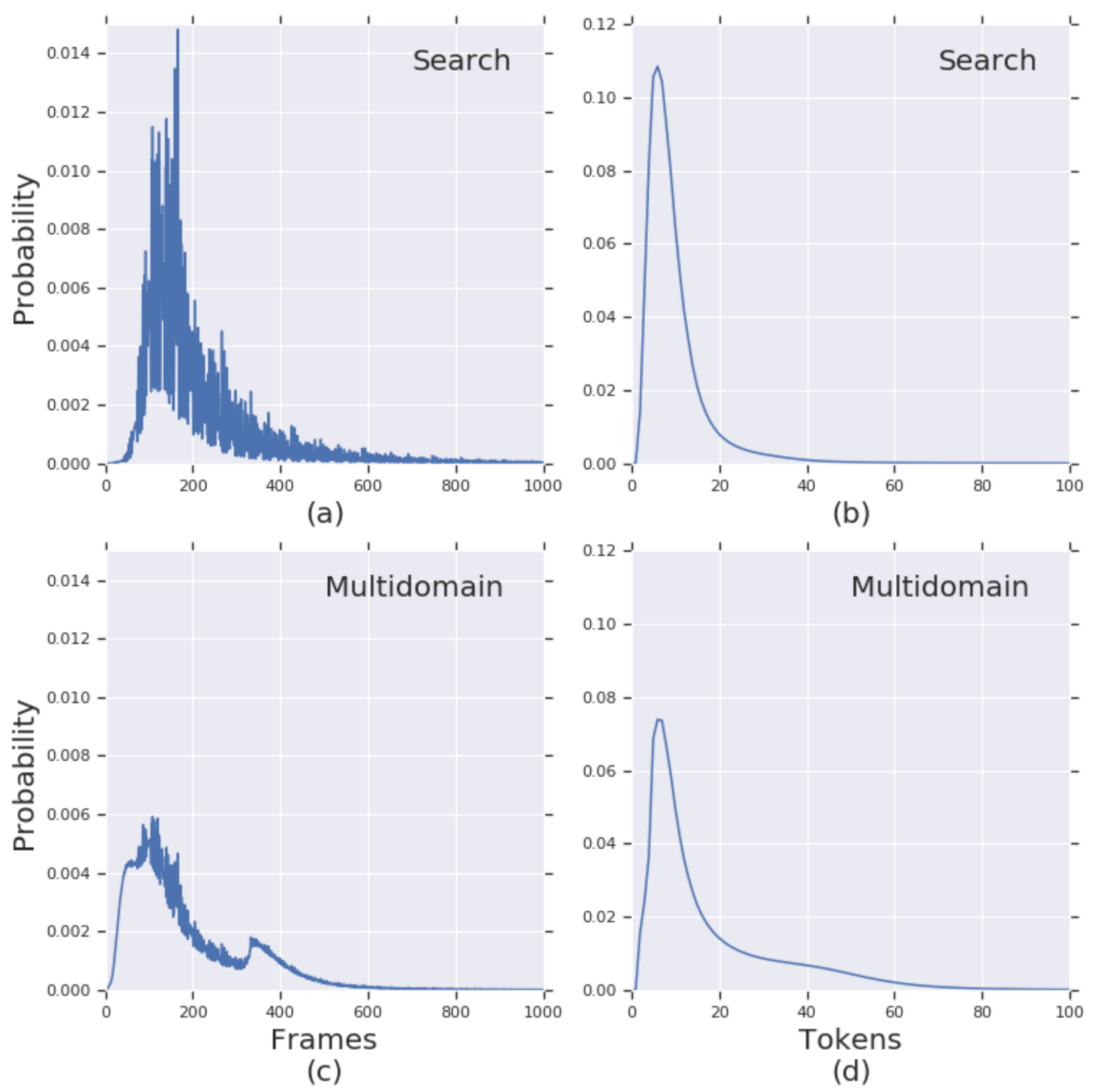}
  \caption{Distribution of number of input frames and output tokens: (a) Number of frames for search subset (b) Number of output tokens for search subset (c) Number of frames for the multidomain set (d) Number of output tokens for the multidomain set.}
  \label{fig:uttlengths}
\end{figure}

Fig.~\ref{fig:uttlengths} shows the distribution of the number of frames, and the number of output tokens for the search and multidomain sets.
As can be seen, the search subset has a median length of $8$ output word-pieces; the $99^\text{th}$ percentile token length is $39$ tokens. The multidomain set, however, has more variation: the set has a median length of $11$ tokens and the $99^\text{th}$ percentile token length is 71 tokens.
A similar increase in variance can also be seen in the distribution of the number of input frames.

For evaluation, we use test sets that are representative of the training acoustic conditions, but with varying lengths. 
Accordingly, we define test sets sampled from a distribution similar to the search and telephony training sets.
We also report results on a YouTube test set that is acoustically similar to YouTube training set, but contains much longer utterances.
To further test generalization to long utterance lengths, we create a TTS-Audiobook test set by synthesizing passages containing at least $200$ words from a novel~\cite{Altsheler1909} using a parametric text-to-speech (TTS) system~\cite{gonzalvo2016recent}, with a single voice.
Finally, we report results on a call-center test set, which is acoustically most similar to the telephony training set, but is unsegmented and contains much longer utterances.
In Tab.~\ref{tab:test_data} we show statistics of the various test sets used in our experiments.
\begin{table}[ht]
  \caption{Utterance length statistics for test data in this work.}
  \label{tab:test_data}
  \centering
  \begin{tabular}{crrr}
    \toprule
    \textbf{Test set} & \textbf{Total} & \textbf{Mean}  & \textbf{Median}\\
    \textbf{} & \textbf{(hours)} & \textbf{(sec.)}  & \textbf{(sec.)}\\
    \midrule
    Search        & $60$   & $7.6$   & $5.8$ \\
    TTS-Audiobook & $7$    & $65.6$  & $62.7$ \\
    Telephony     & $10$   & $4.7$   & $3.7$ \\
    YouTube       & $26$   & $319.0$ & $303.8$ \\
    Call-center   & $26$   & $220.4$ & $133.6$ \\
    \bottomrule
  \end{tabular}
\end{table}

\section{Results}
\label{sec:results}

\subsection{Generalization of Search Models on Long-Form Sets}
\label{sec:generalization-to-long-form}
We report results of the model trained only on the search subset in Tab.~\ref{tab:seq}, where we also indicate deletion (D), insertion (I) and substitution (S) rates.
The table highlights the generalization problems discussed in the earlier sections. 
As can be seen, the model performs well on the in-domain search test set, obtaining a WER of 4.9\%.
The model is trained with a mix of 8 kHz and 16 kHz data, which helps with the performance on the telephony test set, which is at 17.5\%, but performance on the other sets is poor.
On the YouTube set, this model achieves a WER of 69.1\%.
On the acoustically simpler TTS-Audiobook set, the model obtains a WER of 49.0\%.
These results clearly indicate that the model trained only on a single domain with unique acoustic characteristics, fails to generalize to new conditions. 
For TTS-Audiobook, YouTube and call-center test sets, the deletion rates are markedly high. 
An analysis of the results indicates that for these acoustically distinct sets, with significantly longer utterances, the model prefers predicting sequences of blanks instead of valid words.
\begin{table}[ht]
  \caption{Results using models trained on search data. {D~/~I~/~S} stands for deletions, insertions and substitutions.}
  \label{tab:seq}
  \centering
  \begin{tabular}{lcrr}
    \toprule
    {\textbf{Test set}} & {\textbf{Sample rate}} & {\textbf{WER (\%)}} & {\textbf{D~/~I~/~S (\%)}} \\
    \midrule
    Search        & {16 kHz} & $4.9$  & $1.0~/~0.8~/~3.2$  \\
    TTS-Audiobook & {16 kHz} & $49.0$ & $43.6~/~0.5~/~4.9$ \\
    Telephony     & {~8 kHz} & $17.5$ & $8.6~/~1.6~/~7.4$ \\
    YouTube       & {16 kHz} & $69.1$ & $65.4~/~0.5~/~3.2$ \\
    Call-center   & {~8 kHz} & $58.0$ & $52.2~/~1.0~/~4.8$ \\
    \bottomrule
  \end{tabular}
\end{table}

\subsection{Multidomain Training: Impact of Data Diversity}
\label{subsec:seq_vs_md} 
Next, we compare how increasing data diversity helps improve performance.
An important consideration when using multidomain data is determining how utterances should be sampled from each of these domains during training.
This is particularly important, since different domains might have different amounts of data.
In previous work~\cite{ghahremani17-transfer}, Ghahremani \textit{et al.} recommend that gradients of utterances from a particular domain should be scaled by the inverse of the square root of the number of utterances in the domain, thus effectively over-sampling domains with less data.
In Tab.~\ref{tab:md} we compare three sampling strategies: {1)~Sample} from each domain in Tab.~\ref{tab:train_data} with equal probability (Uniform-Domain); {2)~Further} divide each domain into subdomains\footnote{For example, YouTube domain is divided into subdomains based on content, like News, Education, etc.}, and sample from each subdomain with equal probability (Uniform-Subdomain); {3)~Sample} from each domain with probability proportional to the total number of utterances in the domain (Count-Weighted). 
As can be seen in the table, for E2E ASR models, we find that, contrary to~\cite{ghahremani17-transfer}, the best strategy is to sample utterances proportional to the amount of training data in each domain.
In fact, if we over-sample domains or sub-domains with less data, the model rapidly overfits on these domains.
\begin{table}[ht]
  \caption[The LOF Caption]{Comparison of various sampling strategies when training with multidomain data.\footnotemark}
  \label{tab:md}
  \centering
  \begin{tabular}{lrrr}
    \toprule
    {\textbf{Test set}} & \multicolumn{3}{c}{\textbf{WER (\%)}} \\
    {} & {\textbf{Uniform-}} & {\textbf{Uniform-}} & {\textbf{Count-}} \\
    {} & {\textbf{Subdomain}} & {\textbf{Domain}} & {\textbf{Weighted}} \\
    \midrule
    Search    & $5.7$  & $5.1$  & $4.9$  \\
    Telephony & $13.3$ & $11.9$ & $8.7$  \\
    YouTube   & $14.6$ & $12.7$ & $11.9$ \\
    \bottomrule
  \end{tabular}
\end{table}
\footnotetext{These results are obtained using a model with a different word-piece vocabulary, and are not directly comparable with the other results in the paper.}

In Tab.~\ref{tab:seq_vs_md} we compare WERs using models trained on search subset, and the multidomain training set using the Count-Weighted strategy.
As can be seen in the table, increasing training data diversity significantly improves performance in a range of conditions. 
Even on the search test set, we see a 6\% relative WER reduction. 
Performance on the TTS-Audiobook improves dramatically from 49.0\% to 4.7\%.
Similar gains are seen in telephony and the YouTube sets.
Interestingly, on the harder call-center test set, multidomain training helps improve performance by 40\% relative.
From the deletion, insertion and substitution rates shown in the table, it can be seen that the multidomain model has much fewer deletions than the model trained with the search subset. The multidomain model still has high deletions for the call-center test set, likely because of the differences in acoustics compared to the training data -- the utterances in this set are typically long and have much longer pauses.
\begin{table}[ht]
  \caption{Results comparing models trained on search and multidomain data. For the multidomain model, (deletions / insertions / substitutions) are also shown.}
  \label{tab:seq_vs_md}
  \centering
  \begin{tabular}{lrr}
    \toprule
    {\textbf{Test set}} & \multicolumn{2}{c}{\textbf{WER (\%) (D~/~I~/~S)}} \\
    {} & {\textbf{Search}} & {\textbf{Multidomain}} \\
    \midrule
    Search        & $4.9$   & $4.6~(0.8~/~0.8~/~3.0)$  \\
    TTS-Audiobook & $49.0$  & $4.7~(0.5~/~0.5~/~3.7)$  \\
    Telephony     & $17.5$  & $7.9~(1.5~/~2.3~/~4.0)$  \\
    YouTube       & $69.1$  & $12.0~(3.1~/~1.5~/~7.3)$ \\
    Call-center   & $58.0$  & $34.7~(26.5~/~2.4~/~5.9)$ \\
    \bottomrule
  \end{tabular}
\end{table}

\subsection{Simulating Long-Form Utterances During Training}
Finally, we present results using random state sampling and random state passing techniques. For RSS, we only sample initial states for the encoder LSTMs, since we empirically found it to work better than sampling initial states of both encoder and projection network.
In Tab.~\ref{tab:seq_state} we report results when only the search subset is used for training.
As can be seen, on the search test set, the baseline model, which uses all-0 initial LSTM states, and both RSS and RSP perform similarly, obtaining WERs of 4.9\%, 4.8\%, and 5.0\%, respectively.
On the TTS-Audiobook set, RSS improves WER by 37\% relative, and RSP improves WER by 67\% relative. 
While RSP consistently improves performance, it alone is insufficient to completely account for acoustic dissimilarities between training and testing: For example, on the YouTube test set, even though RSP improves WERs by 11\% relative, the overall WER is still as high as 61.7\%. The effect of state-passing is similar to multidomain training; on the TTS-Audiobook set, the deletion rates were reduced from 43.6 percentage points to 6.9.
\begin{table}[ht]
  \caption{Results using models trained on search, with various LSTM state initialization. Baseline uses the typical zero-state initialization, RSS and RSP stand for random state sampling and random state passing, respectively.}
  \label{tab:seq_state}
  \centering
  \begin{tabular}{lrrr}
    \toprule
    {\textbf{Test set}} & \multicolumn{3}{c}{\textbf{WER (\%)}} \\
    {} & {\textbf{Baseline}} & {\textbf{RSS}} & {\textbf{RSP}} \\
    \midrule
    Search        & $4.9$   & $4.8$  & $5.0$  \\
    TTS-Audiobook & $49.0$  & $30.8$ & $16.2$ \\
    Telephony     & $17.5$  & $17.3$ & $17.1$ \\
    YouTube       & $69.1$  & $66.1$ & $61.7$ \\
    Call-center   & $58.0$  & $60.8$ & $53.9$ \\ 
    \bottomrule
  \end{tabular}
\end{table}
\begin{figure*}[htb]
  \centering
  \includegraphics[width=0.9\linewidth]{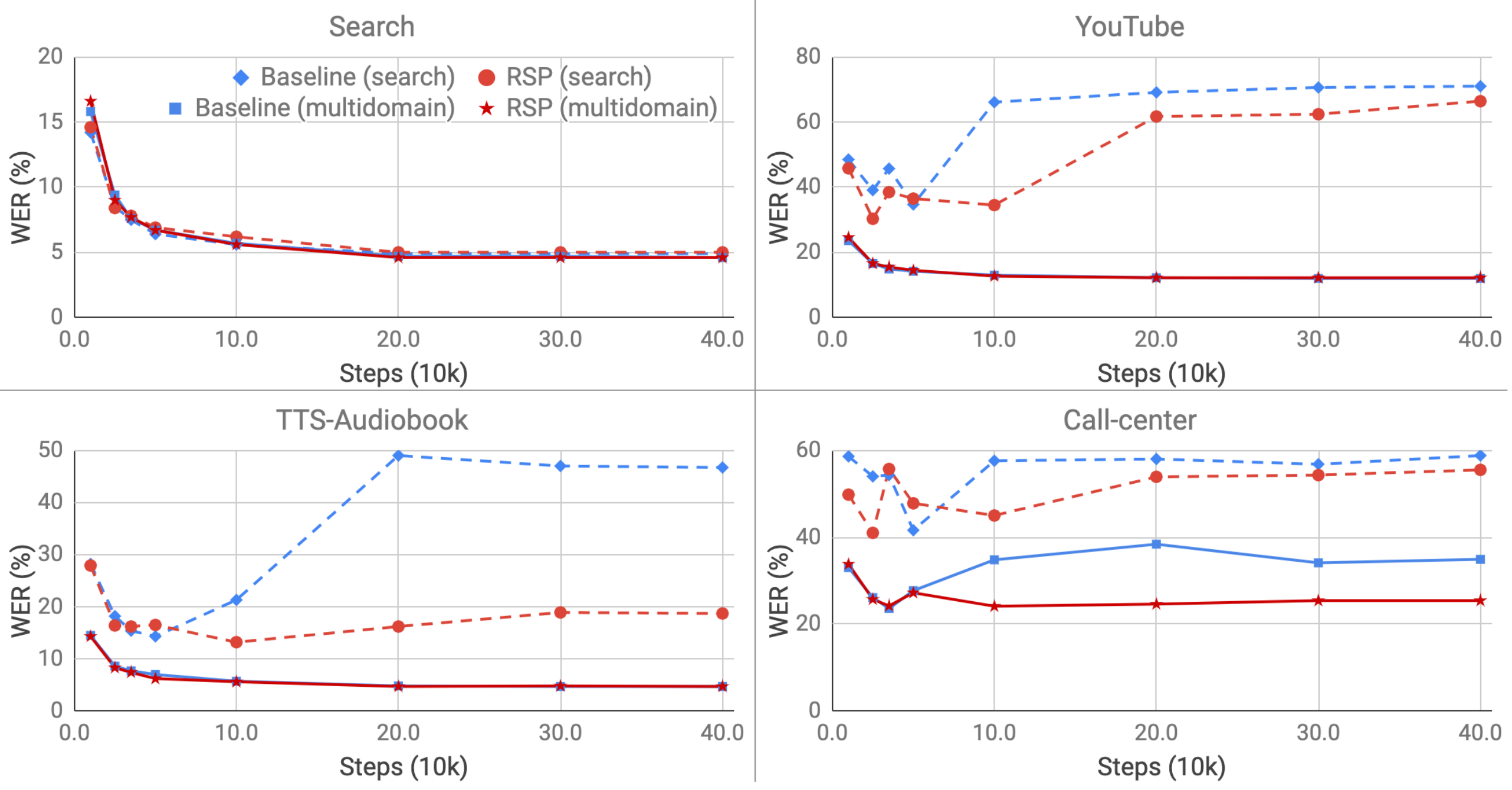}
  \caption{WER as a function of training epoch, when using the model trained on search and multidomain data, with and without random state passing. Results are shown on 4 different test sets -- Search, TTS-Audiobook, YouTube and Call-center.}
  \label{fig:train_vs_epoch}
\end{figure*}

Tab.~\ref{tab:md_state} shows results when the multidomain set is used for training.
As can be seen in the table, RSS and RSP perform comparably with the baseline on the in-domain test sets.
However, on the unseen domain of call-center utterances, these techniques significantly improve performance over the baseline: RSS improves WER by 8\% relative, and RSP improves it by 27\% relative. Compared to the baseline model, RSP reduces the deletion rate from 26.5\% to 17.1\%.
When combined with multidomain training, RSS and RSP can improve performance further and are thus complementary.
\begin{table}[ht]
  \caption{Results using models trained on multidomain data, with various LSTM state initialization. Baseline uses the typical zero-state initialization, RSS and RSP stand for random state sampling and random state passing, respectively.}
  \label{tab:md_state}
  \centering
  \begin{tabular}{lrrr}
    \toprule
    {\textbf{Test set}} & \multicolumn{3}{c}{\textbf{WER (\%)}} \\
    {} & {\textbf{Baseline}} & {\textbf{RSS}} & {\textbf{RSP}} \\
    \midrule
    Search        & $4.6$  & $4.5$  & $4.6$  \\
    TTS-Audiobook & $4.7$  & $4.6$  & $4.7$  \\
    Telephony     & $7.9$  & $8.1$  & $7.9$  \\
    YouTube       & $12.0$ & $11.9$ & $12.2$ \\
    Call-center   & $34.7$ & $32.1$ & $25.4$ \\ 
    \bottomrule
  \end{tabular}
\end{table}

\section{Analysis: Generalization During Training}
\label{sec:analysis}
In the previous section, we demonstrated that increasing training data diversity and simulating long-form utterances can significantly improve performance. 
In this section, we analyze how performance varies as training progresses, and how continued training affects generalization as the model learns to fit the training data better. 
We investigate generalization performance of the models trained on search and multidomain data, on both in-domain (Search and YouTube) and out-of-domain test sets (TTS-Audiobook and Call-center). 
These results are presented in Fig.~\ref{fig:train_vs_epoch}, where we contrast the baseline  model (blue), and the same model trained with RSP (red) using search (solid line) and multidomain (dashed line) training sets. 

As can be seen in the figure, the models trained only on search data perform similarly on the in-domain search test set, with performance continuing to improve as training proceeds.
However, when examining performance on the remaining test sets, it is apparent that generalization performance on unseen domains continues to degrade as performance on the in-domain test set continues to improve.
It is also interesting to note that on the TTS-Audiobook test set -- which is not as acoustically challenging as YouTube -- RSP can significantly stabilize performance when recognizing long-form audio, although performance does begin to degrade slightly after 100k steps, which translates to approximately 13 epochs.
On the more challenging YouTube test set, which contains longer audio and more acoustic diversity, the WER of the RSP model continues to severely degrade after about 100k steps. 
Interestingly, after 50k steps the WER on the YouTube test set is 34.5\%, which is significantly better than the model at convergence.
Although RSP addresses some forms of overfitting, as shown by the results on the TTS-Audiobook set, it fails to prevent regression in performance on the YouTube set which, apart from length mismatch, has significant acoustic dissimilarities with the search training set.

The multidomain-trained models show similar trends:
On the search, TTS-Audiobook and YouTube test sets, both the baseline and RSP models get better as training progresses.
On the Call-center test set, however, the RSP model degrades much slower than the baseline, achieving significantly better performance at convergence.

Comparing the models trained on search versus the ones trained on multidomain data, we can see that training with a variety of data prevents the model from significantly regressing on all of the diverse test conditions.
Even on the Call-center test set, the relative degradation in performance of the multidomain-trained models is much lower than the degradation of the search-trained models.

\section{Conclusions}
\label{sec:conclusions}
In this work, we investigated the generalization abilities of streaming E2E ASR models -- specifically, the RNN-T model -- in order to determine its robustness to domain mismatches between training and inference.
We find that RNN-T models are particularly susceptible to such mismatches, resulting in poor performance on unseen domains.
We also identified a related, but distinct failure mode: the model's inability to recognize long-form speech utterances during inference, when the model is only exposed to short segments during training.
In order to overcome these issues, we proposed and evaluated two methods: increasing training data diversity, and simulating long-form audio during training by manipulating the models internal recurrent state.
In experimental results, we find that both proposed techniques are complementary, resulting in significant performance improvements on unseen domains containing long utterances.

\section{Acknowledgements}
The authors would like to thank Y. He, and A. Tripathi for helpful comments and suggestions; M. Elfeky, H. Liao, H. Soltau, and Y. Shangguan for assistance in creating training sets used in this work; and  D. Zhao for assistance in creating the TTS-Audiobook test set used in this work.

\bibliographystyle{IEEEbib}

\begin{thebibliography}{10}

\bibitem{HintonDengYuDahlEtAl12}
G.~{Hinton}, L.~{Deng}, D.~{Yu}, G.~E. {Dahl}, A.~{Mohamed}, N.~{Jaitly},
  A.~{Senior}, V.~{Vanhoucke}, P.~{Nguyen}, T.~N. {Sainath}, and
  B.~{Kingsbury},
\newblock ``{Deep Neural Networks for Acoustic Modeling in Speech Recognition:
  The Shared Views of Four Research Groups},''
\newblock {\em IEEE Signal Processing Magazine}, vol. 29, no. 6, pp. 82--97,
  Nov 2012.

\bibitem{MorganBourlard95}
N.~{Morgan} and H.~{Bourlard},
\newblock ``{Continuous Speech Recognition},''
\newblock {\em IEEE Signal Processing Magazine}, vol. 12, no. 3, pp. 24--42,
  May 1995.

\bibitem{HannunCaseCasperCatanzaroEtAl14}
A.~Hannun, C.~Case, J.~Casper, B.~Catanzaro, G.~Diamos, E.~Elsen, R.~Prenger,
  S.~Satheesh, S.~Sengupta, A.~Coates, et~al.,
\newblock ``{Deep speech: Scaling Up End-to-End Speech Recognition},''
\newblock {\em arXiv preprint arXiv:1412.5567}, 2014.

\bibitem{AmodeiAnathanarayananAnubhaiBaiEtAl16}
D.~Amodei, S.~Ananthanarayanan, R.~Anubhai, J.~Bai, E.~Battenberg, C.~Case,
  J.~Casper, B.~Catanzaro, Q.~Cheng, G.~Chen, et~al.,
\newblock ``{Deep Speech 2: End-to-End Speech Recognition in English and
  Mandarin},''
\newblock in {\em Proc. of ICML}, 2016.

\bibitem{soltau2016neural}
H.~Soltau, H.~Liao, and H.~Sak,
\newblock ``{Neural Speech Recognizer: Acoustic-to-Word {LSTM} Model for Large
  Vocabulary Speech Recognition},''
\newblock in {\em Proc. of Interspeech}, 2017.

\bibitem{AudhkhasiRamabhadranSaonPichenyEtAl17}
K.~Audhkhasi, B.~Ramabhadran, G.~Saon, M.~Picheny, and D.~Nahamoo,
\newblock ``{Direct Acoustics-to-Word Models for English Conversational Speech
  Recognition},''
\newblock in {\em Proc. of Interspeech}, 2017.

\bibitem{Graves12}
A.~{Graves},
\newblock ``{Sequence Transduction with Recurrent Neural Networks},''
\newblock {\em arXiv preprint arXiv:1211.3711}, 2012.

\bibitem{GravesMohamedHinton13}
A.~{Graves}, A.~{Mohamed}, and G.~{Hinton},
\newblock ``{Speech Recognition with Deep Recurrent Neural Networks},''
\newblock in {\em Proc. of ICASSP}, 2013.

\bibitem{ChorowskiBahdanauSerdyukChoEtAl15}
J.~{Chorowski}, D.~{Bahdanau}, D.~{Serdyuk}, K.~{Cho}, and Y.~{Bengio},
\newblock ``{Attention-Based Models for Speech Recognition},''
\newblock in {\em Proc. of NIPS}, 2015.

\bibitem{ChanJaitlyLeVinyals16}
W.~{Chan}, N.~{Jaitly}, Q.~V. {Le}, and O.~{Vinyals},
\newblock ``{Listen, Attend and Spell: A Neural Network for Large Vocabulary
  Conversational Speech Recognition},''
\newblock in {\em Proc. of ICASSP}, 2016.

\bibitem{PrabhavalkarRaoSainathLiEtAl17}
R.~Prabhavalkar, K.~Rao, T.~N. Sainath, B.~Li, L.~Johnson, and N.~Jaitly,
\newblock ``{A Comparison of Sequence-to-Sequence Models for Speech
  Recognition},''
\newblock in {\em Proc. of Interspeech}, 2017.

\bibitem{BattenbergChenChildCoatesEtAl17}
E.~{Battenberg}, J.~{Chen}, R.~{Child}, A.~{Coates}, Y.~G.~Y. {Li}, H.~{Liu},
  S.~{Satheesh}, A.~{Sriram}, and Z.~{Zhu},
\newblock ``{Exploring Neural Transducers for End-to-end Speech Recognition},''
\newblock in {\em Proc. of ASRU}, 2017.

\bibitem{HoriWatanabeZhangChan2017}
T.~Hori, S.~Watanabe, Y.~Zhang, and W.~Chan,
\newblock ``{Advances in Joint CTC-Attention Based End-to-End Speech
  Recognition with a Deep CNN Encoder and RNN-LM},''
\newblock in {\em Proc. of Interspeech}, 2017.

\bibitem{ZeyerIrieSchluterNey2018}
A.~Zeyer, K.~Irie, R.~Schl{\"u}ter, and H.~Ney,
\newblock ``{Improved Training of End-to-end Attention Models for Speech
  Recognition},''
\newblock in {\em Proc. of Interspeech}, 2018.

\bibitem{ParkChanZhangChiuEtAl19}
D.~S. {Park}, W.~{Chan}, Y.~{Zhang}, C.-C. {Chiu}, B.~{Zoph}, E.~D. {Cubuk},
  and Q.~V. {Le},
\newblock ``{{SpecAugment}: A Simple Data Augmentation Method for Automatic
  Speech Recognition},''
\newblock {\em arXiv preprint arXiv:1904.08779}, 2019.

\bibitem{RaoSakPrabhavalkar17}
K.~{Rao}, H.~{Sak}, and R.~{Prabhavalkar},
\newblock ``{Exploring Architectures, Data and Units for Streaming End-to-End
  Speech Recognition with RNN-Transducer},''
\newblock in {\em Proc. of ASRU}, 2017.

\bibitem{ChiuSainathWuPrabhavalkarEtAl18}
C.-C. {Chiu}, T.~N. {Sainath}, Y.~{Wu}, R.~{Prabhavalkar}, P.~{Nguyen},
  Z.~{Chen}, A.~{Kannan}, R.~J. {Weiss}, K.~{Rao}, E.~{Gonina}, N.~{Jaitly},
  B.~{Li}, J.~{Chorowski}, and M.~{Bacchiani},
\newblock ``{State-of-the-Art Speech Recognition with Sequence-to-Sequence
  Models},''
\newblock in {\em Proc. of ICASSP}, 2018.

\bibitem{TaraRuoming19}
T.~N. Sainath, R.~Pang, D.~Rybach, Y.~He, R.~Prabhavalkar, W.~Li, M.~Visontai,
  Q.~Liang, T.~Strohman, Y.~Wu, I.~McGraw, and C.-C. Chiu,
\newblock ``{{Two-Pass End-to-End Speech Recognition}},''
\newblock in {\em Proc. of Interspeech}, 2019.

\bibitem{NarayananMisraSimPundakEtAl18}
A.~{Narayanan}, A.~{Misra}, K.~C. {Sim}, G.~{Pundak}, A.~{Tripathi},
  M.~{Elfeky}, P.~{Haghani}, T.~{Strohman}, and M.~{Bacchiani},
\newblock ``{Toward Domain-Invariant Speech Recognition via Large Scale
  Training},''
\newblock in {\em Proc. of SLT}, 2018.

\bibitem{jaitly2015neural}
N.~Jaitly, D.~Sussillo, Q.~V. Le, O.~Vinyals, I.~Sutskever, and S.~Bengio,
\newblock ``{A Neural Transducer},''
\newblock {\em arXiv preprint arXiv:1511.04868}, 2015.

\bibitem{ChiuRaffel18}
C.-C. Chiu and C.~Raffel,
\newblock ``{Monotonic Chunkwise Attention},''
\newblock in {\em Proc. of ICLR}, 2018.

\bibitem{Chiu19longform}
C.-C. Chiu, W.~Han, Y.~Zhang, R.~Pang, S.~Kishchenko, P.~Nguyen, H.~Soltau,
  A.~Narayanan, H.~Liao, S.~Zhang, A.~Kannan, R.~Prabhavalkar, Z.~Chen,
  T.~Sainath, and Y.~Wu,
\newblock ``{A Comparison of End-to-end Models for Long-form Speech
  Recognition},''
\newblock in {\em Proc. of ASRU}, 2019.

\bibitem{SchusterNakajima12}
M.~{Schuster} and K.~{Nakajima},
\newblock ``{Japanese and Korean Voice Search},''
\newblock in {\em Proc. of ICASSP}, 2012.

\bibitem{GravesFernandezGomezSchmidhuber06}
A.~Graves, S.~Fern{\'a}ndez, F.~Gomez, and J.~Schmidhuber,
\newblock ``{Connectionist Temporal Classification: Labelling Unsegmented
  Sequence Data with Recurrent Neural Networks},''
\newblock in {\em Proc. of ICML}, 2006.

\bibitem{HochreiterSchmidhuber97}
S.~Hochreiter and J.~Schmidhuber,
\newblock ``{Long Short-Term Memory},''
\newblock {\em Neural Computation}, vol. 9, no. 8, pp. 1735--1780, 1997.

\bibitem{SutskeverVinyalsLe14}
I.~Sutskever, O.~Vinyals, and Q.~V. Le,
\newblock ``{Sequence to Sequence Learning with Neural Networks},''
\newblock in {\em Proc. of NIPS}, 2014.

\bibitem{Lowerre76}
B.~T. Lowerre,
\newblock {\em {The Harpy Speech Recognition System.}},
\newblock Ph.D. thesis, Carnegie Mellon University, Pittsburgh, PA, USA, 1976.

\bibitem{hochreiter1998vanishing}
S.~Hochreiter,
\newblock ``{The Vanishing Gradient Problem During Learning Recurrent Neural
  Nets and Problem Solutions},''
\newblock {\em International Journal of Uncertainty, Fuzziness and
  Knowledge-Based Systems}, vol. 6, no. 02, pp. 107--116, 1998.

\bibitem{peddinti2016mtr}
V.~Peddinti, V.~Manohar, Y.~Wang, D.~Povey, and S.~Khudanpur,
\newblock ``{Far-Field {ASR} Without Parallel Data.},''
\newblock in {\em Proc. of Interspeech}, 2016.

\bibitem{kim2017mtr}
C.~Kim, A.~Misra, K.~Chin, T.~Hughes, A.~Narayanan, T.~N. Sainath, and
  M.~Bacchiani,
\newblock ``{Generation of Large-Scale Simulated Utterances in Virtual Rooms to
  Train Deep-Neural Networks for Far-Field Speech Recognition in {Google
  Home}},''
\newblock in {\em Proc. of Interspeech}, 2017.

\bibitem{ghahremani17-transfer}
P.~Ghahremani, V.~Manohar, H.~Hadian, D.~Povey, and S.~Khudanpur,
\newblock ``{Investigation of Transfer Learning for {ASR} Using {LF-MMI}
  Trained Neural Networks},''
\newblock in {\em Proc. of ASRU}, 2017.

\bibitem{Werbos88}
P.~J. Werbos,
\newblock ``{Generalization of Backpropagation with Application to a Recurrent
  Gas Market Model},''
\newblock {\em {Neural Networks}}, vol. 1, no. 4, pp. 339--356, 1988.

\bibitem{DaiYangYangCarbonnelEtAl19}
Z.~{Dai}, Z.~{Yang}, Y.~{Yang}, J.~{Carbonell}, Q.~V. {Le}, and
  R.~{Salakhutdinov},
\newblock ``{Transformer-XL: Attentive Language Models Beyond a Fixed-Length
  Context},''
\newblock {\em arXiv preprint arXiv:1901.02860}, 2019.

\bibitem{Yu2013FeatureLearningDNN}
D.~Yu, M.~L. Seltzer, J.~Li, J.-T. Huang, and F.~Seide,
\newblock ``{Feature Learning in Deep Neural Networks - Studies on Speech
  Recognition Tasks},''
\newblock in {\em Proc. of ICLR}, 2013.

\bibitem{HeSainathPrabhavalkarMcGrawEtAl19}
Y.~{He}, T.~N. {Sainath}, R.~{Prabhavalkar}, I.~{McGraw}, R.~{Alvarez},
  D.~{Zhao}, D.~{Rybach}, A.~{Kannan}, Y.~{Wu}, R.~{Pang}, Q.~{Liang},
  D.~{Bhatia}, Y.~{Shangguan}, B.~{Li}, G.~{Pundak}, K.~C. {Sim}, T.~{Bagby},
  S.~{Chang}, K.~{Rao}, and A.~{Gruenstein},
\newblock ``{Streaming End-to-End Speech Recognition for Mobile Devices},''
\newblock in {\em Proc. of ICASSP}, 2019.

\bibitem{abadi2016tensorflow}
M.~Abadi, P.~Barham, J.~Chen, Z.~Chen, A.~Davis, J.~Dean, M.~Devin,
  S.~Ghemawat, G.~Irving, M.~Isard, et~al.,
\newblock ``{Tensorflow: A System for Large-scale Machine Learning},''
\newblock in {\em Proc. of OSDI}, 2016.

\bibitem{ShenNguyenWuChenEtAl19}
J.~{Shen}, P.~{Nguyen}, Y.~{Wu}, Z.~{Chen}, M.~X. {Chen}, Y.~{Jia},
  A.~{Kannan}, T.~{Sainath}, Y.~{Cao}, C.-C. {Chiu}, et~al.,
\newblock ``{Lingvo: a Modular and Scalable Framework for Sequence-to-Sequence
  Modeling},''
\newblock {\em arXiv preprint arXiv:1902.08295}, 2019.

\bibitem{jouppi2017datacenter}
N.~P. Jouppi, C.~Young, N.~Patil, D.~Patterson, G.~Agrawal, R.~Bajwa, S.~Bates,
  S.~Bhatia, N.~Boden, Al~Borchers, et~al.,
\newblock ``{In-datacenter Performance Analysis of a Tensor Processing Unit},''
\newblock in {\em Proc. of Annual International Symposium on Computer
  Architecture (ISCA)}. IEEE, 2017.

\bibitem{kingma2014adam}
D.~P. Kingma and J.~Ba,
\newblock ``{Adam: A method for Stochastic Optimization},''
\newblock {\em arXiv preprint arXiv:1412.6980}, 2014.

\bibitem{liao2013large}
H.~Liao, E.~McDermott, and A.~Senior,
\newblock ``{Large Scale Deep Neural Network Acoustic Modeling with
  Semi-supervised Training Data for YouTube Video Transcription},''
\newblock in {\em Proc. of ASRU}. IEEE, 2013.

\bibitem{Altsheler1909}
J.~A. Altsheler,
\newblock {\em The Last of the Chiefs: A Story of the Great Sioux War},
\newblock Grosset \& Dunlap, 1909.

\bibitem{gonzalvo2016recent}
X.~Gonzalvo, S.~Tazari, C.-A. Chan, M.~Becker, A.~Gutkin, and H.~Silen,
\newblock ``{Recent Advances in Google Real-time HMM-driven Unit Selection
  Synthesizer},''
\newblock in {\em Proc. of Interspeech}, 2016.

\end{thebibliography}
\label{sec:ref}

\end{document}